\documentclass[11pt,]{article}
\usepackage[]{mathpazo}
\usepackage{amssymb,amsmath}
\usepackage{ifxetex,ifluatex}
\usepackage{fixltx2e} 
\ifnum 0\ifxetex 1\fi\ifluatex 1\fi=0 
  \usepackage[T1]{fontenc}
  \usepackage[utf8]{inputenc}
\else 
  \ifxetex
    \usepackage{mathspec}
  \else
    \usepackage{fontspec}
  \fi
  \defaultfontfeatures{Ligatures=TeX,Scale=MatchLowercase}
\fi
\IfFileExists{upquote.sty}{\usepackage{upquote}}{}
\IfFileExists{microtype.sty}{%
\usepackage{microtype}
\UseMicrotypeSet[protrusion]{basicmath} 
}{}
\usepackage[margin=1in]{geometry}
\usepackage{hyperref}
\PassOptionsToPackage{usenames,dvipsnames}{color} 
\hypersetup{unicode=true,
            pdftitle={Herding behavior in cryptocurrency markets},
            colorlinks=true,
            linkcolor=Maroon,
            citecolor=Blue,
            urlcolor=blue,
            breaklinks=true}
\usepackage{longtable,booktabs}
\usepackage{graphicx,grffile}
\makeatletter
\def\maxwidth{\ifdim\Gin@nat@width>\linewidth\linewidth\else\Gin@nat@width\fi}
\def\maxheight{\ifdim\Gin@nat@height>\textheight\textheight\else\Gin@nat@height\fi}
\makeatother
\setkeys{Gin}{width=\maxwidth,height=\maxheight,keepaspectratio}
\IfFileExists{parskip.sty}{%
\usepackage{parskip}
}{
\setlength{\parindent}{0pt}
\setlength{\parskip}{6pt plus 2pt minus 1pt}
}
\ifx\paragraph\undefined\else
\let\oldparagraph\paragraph
\renewcommand{\paragraph}[1]{\oldparagraph{#1}\mbox{}}
\fi
\ifx\subparagraph\undefined\else
\let\oldsubparagraph\subparagraph
\renewcommand{\subparagraph}[1]{\oldsubparagraph{#1}\mbox{}}
\fi

\let\rmarkdownfootnote\footnote%
\def\footnote{\protect\rmarkdownfootnote}

\usepackage{titling}

\newcommand{\subtitle}[1]{
  \posttitle{
    \begin{center}\large#1\end{center}
    }
}

  \title{Herding behavior in cryptocurrency markets}
    \pretitle{\vspace{\droptitle}\centering\huge}
  \posttitle{\par}
  \subtitle{Universitat Autònoma de Barcelona\\
Department of Applied Economics}
  \author{Obryan Poyser Calderón}
    \preauthor{\centering\large\emph}
  \postauthor{\par}
      \predate{\centering\large\emph}
  \postdate{\par}
    \date{WORKING PAPER \textbar{} November 2018}

\usepackage{amsmath}
\usepackage{pdflscape}
\usepackage{setspace}
\usepackage{mathtools}
\usepackage{threeparttable}
\newcommand{\blandscape}{\begin{landscape}}
\newcommand{\elandscape}{\end{landscape}}
\usepackage{multirow}
\usepackage{booktabs,caption}
\usepackage{graphicx}

\begin{document}
\maketitle

\hfill \break
\hfill \break
\hfill \break

\hypertarget{abstract}{%
\section*{Abstract}\label{abstract}}
\addcontentsline{toc}{section}{Abstract}

There are no solid arguments to sustain that digital currencies are the
future of online payments or the disruptive technology that some of its
former participants declared when used to face critiques. This paper
aims to solve the cryptocurrency price determination puzzle from a
behavioral finance perspective and trying to find the parallelism
between the literature on biases present in financial markets that serve
as a starting point to understand crypto-markets. Moreover, it is
suggested that cryptocurrencies' prices are driven by herding, hence
this study test behavioral convergence under the assuption that prices
``as-is'' are the coordination mechanism. For this task, it has been
proposed an empirical herding model based on Chang, Cheng, and Khorana
(\protect\hyperlink{ref-Chang2000}{2000}) methodlogy, and expandind the
model both under asymmetric and symmetric conditions and the existence
of different herding regimes by employing the Markov-Switching approach.

\hfill \break

\textbf{keywords:} Cryptocurrencies, Herding, Speculation, Markov
Switching

\newpage

\hypertarget{introduction}{%
\section{Introduction}\label{introduction}}

The digital economy have been increasing the exposure of state-of-art
ideas, opportunities and changes in economics paradigms. By the same
token, cryptocurrencies as well as Blockchain's technology, and other
potential applications are without a doubt a relevant concept that have
emerged on the ``new economy''. One could assure that most of the
interest on cryptocurrencies was fueled by Bitcoin, the first successful
implementation of a peer to peer network that could serve as a payment
method. The responsiveness from the public that Bitcoin had exposed has
been driven in part for the extreme upswings and downwings in prices,
which has been also illustrated in some degree by other alternative
coins such as Ethereum, Ripple, Tokens or Initial Coin Offering (ICO's).
As portrayed by Poyser (\protect\hyperlink{ref-Poyser2018}{2018}), it is
difficult to align a future in which cryptocurrencies make a significant
economic change under current extreme price movements exhibited without
the existence of salient announcements.

The understanding of crashes in stock markets has been a difficult for
economists for several years. Theoretical foundations in financial
economics rely ultimately on the assumption of efficiency of markets.
Nonetheless, several studies have found empirical evidence that
contrariwise the cornerstone of efficient markets. The behavioral
economics uncover systematic deviations from rationality exposed by
investors, instead individuals are victim of their cognitive biases
leading to the existence of financial market inefficiencies, fragility,
and anomalies. Particularly, crypto-currency markets resembles in great
fashion to the criticisms on financial markets exposed by behavioral
finance advocates.

Studies of behavioral finance aim to explain why investors in stock
market settings act as they do. In this work, it is hypothesized that it
is possible to explain cryptocurrencies market prices' puzzle from a
behavioral finance perspective in which investors' cognitive biases play
a major role to explain the volatility. In this context, this paper
makes a literature revision on empirical and theoretical evidence in
which investors' actions have been proved are not aligned with a
rational benchmark, that can also serve as a parallelism to the
crypto-market problem. Furthermore, this paper seeks as well to shed
light on the price setting puzzle by attributing movements to investors
herding behavior, that is, a collective decision-making process in which
prices ``as is'' are the coordination mechanism to investing decision
making. According to the literature, herding can trigger the formation
of speculative bubbles, thus, the main objective of this chapter is to
study cryptocurrency market under the hypothesis that crypto-investors
have limited resources to process information and weak prior knowledge,
as a consequence they rely on others sources to valuate
cryptocurrencies, which can unchain unexpected results.

The paper is structured as follows: first, it has been contextualized
the problem by comparing cryptocurrency market behavior with past
speculative bubbles. Second, it is defined some of the most relevant
theory on financial economics and the transition to behavioral economics
to find the parallilism between crypto-markets and the evolution of the
literature. Third, it has been provided some of the most common biases
evidenced in financial settings and their relation with the
cryptocurrency's case. In the fourth and fifth sections it is explained
the data and the methodology used for this work. Section six show the
empirical results on herding behavior and concludes in section 7 the
main outcomes to the research.

\hypertarget{from-tulips-to-blockchain}{%
\section{From Tulips to Blockchain}\label{from-tulips-to-blockchain}}

Due to the incentive to generate profits from price differentials,
speculation is ubiquitous to a market economy. Long time ago Fisher
(\protect\hyperlink{ref-fisher1896appreciation}{1896}) mentioned:
\emph{``Every chance for gain is eagerly watched. An active and
intelligent speculation is constantly going on, which, so far as it does
not consist of fictitious and gambling transactions, performs a
well-known and provident function for society.''} Hence, a valid
conjecture is that as long as market economy has been developing,
speculation has been growing as well as a natural mechanism to those
individuals who are willing to use strategical information to their own
good.

However, certain conditions such as high degree of speculation by
misvaluations or delinked relationships between risk and loss, are
associated to market inneficiencies. There are several cases of price
booms in financial and non-financial environments have occurred without
any rational explanation, as a result, economic and financial jargon
have created names like, \emph{``crashes''}, \emph{``bubbles''},
\emph{anomalies}, \emph{``financial crisis''} or \emph{``tulip-mania''}.
The latter name refers to the first documented \emph{``speculative
bubble''} when in the XVII century Tulips' prices increased abruptly.
For instance, an rising in trade of goods inside Netherlands, increasing
value in the national currency, a perception of facing the transition to
a new economy, novel colonial possibilities, and an increasingly
prosperous country, which led to create an atmosphere in which the now
called \emph{``Tulip mania''} (Mackay
\protect\hyperlink{ref-mackay2002extraordinary}{1852}; Sornette
\protect\hyperlink{ref-Sornette2003}{2003}).

According to Mackay
(\protect\hyperlink{ref-mackay2002extraordinary}{1852}), tulips' bulbs
were initially imported at retail from Turkey in the middles of the
sixteenth century. During the first stages of the market build-up, bulbs
sales barely covered production costs (Sornette
\protect\hyperlink{ref-Sornette2003}{2003}), albeit, by the end of
1500's professional cultivators and wealthy marketers started to offer
exotic varieties of bulbs which rapidly gather the attention of rich
people, willing to pay extraordinarily high amounts of money for the
bulbs. Sudenly, tulips became a symbol of wealth and as the rumors
dispersed in the society, other socioeconomic groups such as
middle-class people also realized of the possibility to easily obtain
profits from buying low and selling high in the market. As Sornette
(\protect\hyperlink{ref-Sornette2003}{2003}) mentions, the now named
\emph{``tulip-mania''} was perceived as a \emph{``sure thing''},
suddenly, an atmosphere of euphoria where any hesitation was dismissed,
a complete confidence on the rumors of even higher prices led people to
sell houses, properties in order to invest in this activity. Prices of
rare tulip bulbs escalated at a point where there was no rational
concordance with the price of other goods and services (Sornette
\protect\hyperlink{ref-Sornette2003}{2003}). The process of continuous
increases in prices attracted speculators who started to play with the
information in order to generate market fluctuations with the objective
generate profits from arbitrage activities. Interestingly, \emph{``many
individuals grew suddenly rich. A golden bait hung temptingly out before
the people, and, one after the other, they rushed to the tulip marts,
like flies around a honey-pot''} (Mackay
\protect\hyperlink{ref-mackay2002extraordinary}{1852}). As Sornette
(\protect\hyperlink{ref-Sornette2003}{2003}) mentioned, \emph{``the
conditions now generally associated with the first period of a boom were
all present: an increasing currency, a new economy with novel colonial
possibilities, and an increasingly prosperous country together had
created the optimistic atmosphere in which booms are said to grow''}.
Nonetheless, the ``unexpectable predictable'' ocurred, in February 1637
prices collapsed at a 10\% of the peak values shown months before, and
never rose again. The tulip mania has been since that time portrayed as
the characterization of irrationality in markets.

Even though the tulip-mania was one of the former cases in which markets
results deviates greatly from expected, it has not been as striking and
extended as financial crashes. This is exemplified by the major
historical stock market crash occurred in October of 1929 known as
\emph{``Great Crash''} that also set the precedent for the Great
Depression. Other examples are the \emph{``Black Monday''} denoting
stock market boom ensued on October 19, 1987, or the \emph{``Dot-com''}
bubble occurred in the period 1997-2001. As Garber
(\protect\hyperlink{ref-Garber1990}{1990}) portrayed, these events have
\emph{``emerged from specific speculative episodes have been
sufficiently frequent and important that they underpin a strong current
belief among economists that key capital markets sometimes generate
irrational and inefficient pricing and allocational outcomes''}. Given
that the scope and impact such market collapses is significant, they
remain at the central debate between financial economics theorists, who
try to explain the reason to such exceptional deviations from the
\emph{``fundamental values''} and those who believe in markets driven by
psychological factors.

Historically, several products or services that ended in a crash were in
the beginning exposed as \emph{``disruptive''}, \emph{``new''},
\emph{``innovative''}. Cryptocurrencies have not been different, in
Poyser (\protect\hyperlink{ref-Poyser2018}{2018}) I have briefly
described how people describe them as the precursor of the new form of
performing empowered by the people. Certainly, news media have been
playing a relevant role into forming expectations and increasing the
hype, this by broadcasting insiders and speculators' ideas about new
projects, technological implementations, security advances and future
applications. As a result, there is a lot of noise that oftenly displace
the discussion from serious and tangible projects to marketing littered
implementations.

To some extent it is expected such outcome taken in count that
Blockchain is difficult to understand even for information technology
enthusiasts. As expressed before, any disclosure evoking a ``new
economy'' will always be an allure for people ready to obtain profits,
notably in cryptocurrency markets it is accompanied by a perception of
foolproof investment, a sense of low probability of losing money. As
expected, extreme events such as 2013 Bitcoin's price boom not only led
the genesis of mass hysteria, it also incentivated speculators to play
with the information. Same happened by the end of 2017, when Bitcoin's
price had a run up that caught public attention, increasing the
consensus among economists that it is hard to conciliate high volatility
with a the store of value function that any currency should exhibit
(Also discussed in (Poyser \protect\hyperlink{ref-Poyser2018}{2018})).

In a nutshell, price increases attract large groups of investors, who
believe that they can \emph{``jump into bandwagon''} in order to
generate profits easily, even without figuring out how cryptocurrencies
really works and their potential, though aware that the opportunity cost
of missing out is relative high. By the same token, such enthusiasm has
been promoted by news media, thus, price bid ups, create further price
rises, that is, a self-fulfilling prophecy. Under this situation, it is
likely that cryptocurrencies exhibit characteristics of a speculative
bubble as many others have mentioned, however, it is almost impossible
to predict when it is going to occur.

\hypertarget{theory}{%
\section{Theory}\label{theory}}

\hypertarget{an-intent-to-find-the-parallelism-within-economic-theory}{%
\subsection{An intent to find the parallelism within economic
theory}\label{an-intent-to-find-the-parallelism-within-economic-theory}}

It is been said that theory on financial economics formally started in
the 1900 with Bachelier (\protect\hyperlink{ref-Bachelier1900}{1900}),
who was interested in the application of random motion to explain the
movements of prices of a popular investment tool named ``perpetuity
bond''. In order to explain price dynamics, he implemented the random
walk, that is, a path created by the succession of random uncorrelated
steps in which each move is buffeted by a given equal probability. The
first insight present in Bachelier
(\protect\hyperlink{ref-Bachelier1900}{1900}) in that prices movements
will tend to be on the average (zero) given an equal probability of
going ``up'' or ``down'', therefore the trajectories are neutralized or
canceled. Additionally, the level in which prices fluctuate is named
\emph{``fundamental value''}, and any deviation or fluctuation created
by the forces of the market (active participation from multiple
investors) around this value is governed by the so-called Gaussian
distribution. Hence, any possibility of predicting future values is
impossible and therefore there is no deterministic chance to arbitrage
both for sophisticated and unsophisticated investors. This result is
possible due to the constant feedback dynamics products of the constant
participation, that is, any strategic information in hands of an
investor is quickly recognized and eliminated by others in the market
who analyze prices\footnote{Fama
  (\protect\hyperlink{ref-Fama1965}{1965}) describes accurately what an
  efficient market means by saying: \emph{``An efficient market is
  defined as a market where there are large numbers of rational,
  profit-maximizers actively competing, with each trying to predict
  future market values of individual securities, and where important
  current information is almost freely available to all participants. In
  an efficient market, competition among the many intelligent
  participants leads to a situation where, at any point in time, actual
  prices of individual securities already reflect the effects of
  information based both on events that have already occurred and on
  events which, as of now, the market expects to take place in the
  future. In other words, in an efficient market at any point in time,
  the actual price of a security will be a good estimate of its
  intrinsic value''}}. Up to now, I have described the origin and some
of the cornerstone characteristics of the EMH which had been intensively
improved by other significant contribution of Eugene Fama, Stephen Ross,
Robert Merton, Myron Scholes, William Sharpe, among other whom works set
the basis to the contemporaneous modern finance theory. Another
implication (probably simplification as well) is that any asset has a
fundamental and speculative components. According to the EMH the latter
follow a known probabilistic distribution, and constantly fluctuates
around the fundamental. Nonetheless, this argument has the most
controversial given that hardly explain extreme volatility events such
as speculative bubbles and market crashes.

Several conclusions stems from the statements exposed above, first,
financial prices embed inside the sum of all information publicly
provided over time, hence assets price are always correct, and any
deviation in is only product of market interchange\footnote{In Shiller
  (\protect\hyperlink{ref-Shiller2015}{2015}), the author seriously
  critisized such assumption by saying: \emph{``price may appear to be
  too high or too low at times, but, according to the efficient markets
  theory, this appearance must be an illusion.''}}. Second, it is not
possible to forecast any futher price change, therefore, one could not
systematically beat the market (Read
\protect\hyperlink{ref-Read2012}{2012}).

Evidently, a reasonable question to bear about is if such theory is
indeed a good starting point to describe crypto-markets. In this work,
it has been highly focused on other facts that might reduce
crypto-market ambiguity, hence, we will center the attention to
(exacerbated) speculation and its consequences. Speculation has been a
concern that takes back upon the times of John Maynard Keynes, who
proposed a tax on financial transactions that were excessively
speculative. With the purpose of limits the markets to legitimated
investors and thus mitigate the impact on the economy in case a
potential burst (Keynes \protect\hyperlink{ref-keynes1936general}{1936};
Read \protect\hyperlink{ref-Read2012}{2012}). Speculative bubbles have
been increasing in attention since modern finance cannot explain how
events such as crashes in Black Monday, Dot-Com and the financial crisis
in 2008. On this position, Robert Shiller has assuring that financial
market are driven exclusively by behavioral issues among the
participants (speculative component outbound the fundamental effect).

Summarizing, in this study it has been asked if cryptocurrencies can be
analyzed from the efficient markets markup. A first step is to dispel
and characterize commonalities across the different cryptocurrency, that
is, asking if any given cryptocurrency met all the three functions of
money. It is likely several of them meet the unit of account and medium
of exchange functions, nonetheless it is hard to conciliate the store of
value function given that all of them exhibit great volatility. The
second step is to question if any given CC has a fundamental value.
Devotes might considered Blockchain as itself has a value, but how can
it objectively measured? And if it has fundamental value, why are they
so volatile without any reasonable explanation? For many people CC are
investments, independently if they classified as token, ICO,
currency\ldots{} Therefore, in absence of fundamental or intrinsic value
the speculative component is undoubtly driving prices. In the next
section, I will provide the characteristics that crypto-market has and
how they comply with much of the past arguments in favor of the ``animal
spirits''.

\hypertarget{beliefs-formation-and-biases-present-in-cryptocurrencies-markets}{%
\section{Beliefs formation and biases present in cryptocurrencies
markets}\label{beliefs-formation-and-biases-present-in-cryptocurrencies-markets}}

During the Committee on Banking and Financial services in 1998, Alan
Greenspan exposed his views regarding the conjuncture of financial
markets. He mentioned that human behavior is the main factor that drives
markets, and in spite of corrections there is a constant evolution that
makes behavioral issues pervasively which yields with violent and
unexpected results. Even though this is anecdotic, it has a great coming
from someone who served as Chairman of the Federal Reserve of the United
States for almost 30 years. As it had been discussed before, the EMH had
been the central ideological domain of study among the classical
financial theorists and empiricists until the late seventies, however,
in this case, it has several drawbacks to explain large deviations in
prices such as cryptocurrency phenomena.

A basic tenet of classical economic theory is that investment decisions
reflect agents' rationally expectations, that is, decisions are made
using all available information in an efficient manner. A contrasting
view is that investment are also driven by herd behavior, which weakens
the link between information and market outcomes (Devenow and Welch
\protect\hyperlink{ref-Devenow1996}{1996}; Scharfstein and Stein
\protect\hyperlink{ref-Scharfstein1990}{1990}). In one sense, the EMH
was so successful because it seemed to dispel the previously dominant
notion of an irrational market driven by herds\footnote{Keynes
  (\protect\hyperlink{ref-keynes1936general}{1936}) famous adage was
  that the stock market was mostly a beauty contest in which judges
  picked who they thought other judges would pick, rather than who they
  considered to be most beautiful}. The perceptions of Mackay
(\protect\hyperlink{ref-mackay2002extraordinary}{1852}) and
Kindleberger, Aliber, and Wiley
(\protect\hyperlink{ref-Kindleberger2005}{2005}) is that there was
convincing evidence of ``bubbles'' of mass errors caused by the fickle
nature of herds.

Therefore, what has been happening with cryptocurrencies is closely
related to the criticisms on the rationality of investors. For such
reason, behavioral finance tries to unveil market outcomes under the
existence of a large group of irrational investors by studying
real-world investors' beliefs and valuations. Regarding information
sources, it is particularly interesting that cryptocurrencies advising
is mostly available online, naturally, new (and old) investors are
dependent on information on fairly diversified sources, that is,
individuals interested in cryptocurrencies usually form beliefs and
decisions based on two main sources: news and social media. Nowadays,
many trends start in specific forums, in these spaces users share
impressions about last news and recent issues like unexpected upswings
or downswings in cryptocurrencies price or innovations in Blockchain
platform. That is case of Reddit, a social news aggregation website in
which people discuss a wide range of topics, particularly the community
of cryptocurrencies is the biggest one among the internet, with more
than 600.000 subscribers. There are a wide variety of users
(sophisticated and unsophisticated), as a result, opinion formation on
this community unveil different investment strategies such as
discovering a new altercoin or ``smartly'' recognize price
patterns.\footnote{Another type of feedback formation occurs in
  specialized websites that impulse new users to follow
  \emph{``experienced''}, \emph{``professional''} and
  \emph{``successful''} investors, thus, disregarding private
  information, and following others' actions is precisely a clear
  contradiction with the EMH that states randomization irrational
  investors' decisions.}

Expectations formation on other's investor's opinions has been widely
studied for years, for instance Keynes
(\protect\hyperlink{ref-keynes1936general}{1936}) wrote a clever
metaphor to describe the heuristics' individuals performed to invest in
stock markets and newspapers competition for the most beautiful women
among many options during the thirties:

\emph{``\ldots{}so that each competitor has to pick, not those faces
which he himself finds prettiest, but those which he thinks likeliest to
catch the fancy of the other competitors, all of whom are looking at the
problem from the same point of view. It is not a case of choosing those
which, to the best of one's judgment, are really the prettiest, nor even
those which average opinion genuinely thinks the prettiest. We have
reached the third degree where we devote our intelligence to
anticipating what average opinion expects the average opinion to be. And
there are some, I believe, who practice the fourth, fifth and higher
degrees.''}

The scenario described by Keynes\footnote{Another concept attributed to
  Keynes is \emph{``animal spirits''} which originally described
  business calculation, which he considered the role of confidence,
  uncertainty and framing on investment heuristics is inexorable due to
  our human nature, more precisely of \emph{``\ldots{}a spontaneous urge
  to action rather than inaction, and not as the outcome of a weighted
  average of quantitative benefits multiplied by quantitative
  probabilities.''}} seemly relates to cryptocurrencies market, both on
price determination and which of them to choose invest. Discerning the
degree of compliance within the community is a challenging task since
experienced users can take advantage of curious or ignorant ones in
diverse settings, an aspect that I will consider further. Nowadays we
count on more assertive evidence on Keynes' anecdotal arguments thanks
to empirical and experimental evidence (Kahneman and Riepe
\protect\hyperlink{ref-Kahneman1998}{1998}). Until now, I have described
some of the belief formations on feedback, which has been widely studied
behavioral finance literature, thus, it seems that this field is a good
fit to describe Bitcoin market since efficiency is hardly possible for
the existence of many contradictions with the statements of the
Efficient Market Hypothesis.

It is relevant to describe common biases in judgments and
decision-making, also identified as cognitive illusions that people
usually reflect (Kahneman and Riepe
\protect\hyperlink{ref-Kahneman1998}{1998}). It is also related to the
bounded rationality concept attributed to Simon
(\protect\hyperlink{ref-simon1982models}{1982}). He was concerned with
the human decision making \emph{``shortcuts''} that could lead to
suboptimal outcomes. Naturally, there is a vast set of systematic
behavioral biases that characterize individuals in financial-like
markets such as crypto-markets, however, they emerge from a setting in
which heuristics are altered by market participants and diverse signal
and the noise. Moreover, it is been proved that in asset markets the
existence of irrational investors generates deviations from
fundamentals, hence, under the special case with cryptocurrencies the
absence of a parameter of value creates a different puzzle. At this
point, it is relevant to classify the different cognitive biases found
in the literature on which people are affected. Hence, this study aim to
create a properly standardized aggregation based on literature reviews
studies on behavioral finance from (Kahneman and Riepe
\protect\hyperlink{ref-Kahneman1998}{1998}; Kumar and Goyal
\protect\hyperlink{ref-Kumar2015}{2015}; Shiller
\protect\hyperlink{ref-Shiller1999}{1999}; Stracca
\protect\hyperlink{ref-Stracca2004}{2004}; Subrahmanyam
\protect\hyperlink{ref-Subrahmanyam2008}{2008}).

A crucial starting point in decision-making framework is to distinguish
between beliefs and preferences. Beliefs are salient in expectation
formation, and usually, people develop non-optimal judgments in what to
believe due to a set of experimentally proved systematic errors called
biases Kahneman and Riepe (\protect\hyperlink{ref-Kahneman1998}{1998}).
My own view is that people involved in cryptocurrencies market
presumably suffer from several of the same judgment biases that have
been documented in financial markets settings, which can even get
intensified by crypto-market idiosyncratic uncertainty and complexity.

\hypertarget{overconfidence-and-optimism}{%
\subsection{Overconfidence and
optimism}\label{overconfidence-and-optimism}}

Among the biases people display in financial market exist the
exacerbated trust on our own ability, knowledge, and skills, are
entitled as overconfidence which is intrinsically related with optimism.
Moreover, this self-reliance on personal judgments entails concepts such
as miscalibration, over-precision, which are at the same time associated
with an overreaction to random events (Barber and Odean
\protect\hyperlink{ref-Barber2013}{2013}; Barberis and Thaler
\protect\hyperlink{ref-Barberis2002}{2002}; Kahneman and Riepe
\protect\hyperlink{ref-Kahneman1998}{1998}). A classic illustration of
overconfidence bias is the ``better than the average'' beliefs, which is
the perception of a more than proportional of a group's composition that
they perform better than the mean for the same group for certain
activities. For instance, Svenson
(\protect\hyperlink{ref-Svenson1981}{1981}) found 90\% of Swedish car
drivers considered themselves better than the average. Another example
is the seemly narrow uncertainty in judgments, in other words, people
assign significantly less weight on chances of surprises than they
really occur. The typical example is when people were asked to evaluate
1 and 99 percentiles of an index such as exchange rates a year from the
reference point, the resulting 98\% percent confidence interval captured
far less the expected value in comparison with expected intervals
(Alpert and Raiffa \protect\hyperlink{ref-Alpert1982}{1982}). Hence, it
has been proved that uncertainty is considerable high, in fact, the
surprise rate is about 20\% where the accurate calibration would yield
2\%. Also, Barber and Odean (\protect\hyperlink{ref-Barber2001}{2001})
found on the premise that men are more prone to overconfidence than
women, that the former gender trade more and display lower return than
women. Other great contributions can be found in Daniel, Hirshleifer,
and Subrahmanyam (\protect\hyperlink{ref-Daniel1998}{1998}) and Daniel
and Hirshleifer (\protect\hyperlink{ref-Daniel2015}{2015}), which also
stated how unlikely is the purely rational model to explain variability
in stock prices due to systematic departures from rational behavior.

\hypertarget{information-and-social-wisdom}{%
\subsection{Information and social
wisdom}\label{information-and-social-wisdom}}

Confrontation of different ideas has been playing a essential role in
the development of the society. The invention of the printing press is
one of the most significant, if not the most dramatic event that yields
to the conception of information as a near public good. There is
evidence that the decline in the cost of dissemination of knowledge and
ideas due to press accounted for 18 and 68 percent of European city
growth between 1500 and 1600 (Dittmar
\protect\hyperlink{ref-Dittmar2011}{2011}). Nowadays, in the digital
economy, information is no longer a scarce commodity, in fact, the
opposite is exposed, there is an overload of data that demands the
creation of mechanisms to discern which is relevant and which is not. On
this matter, H. Simon accurately described the situation by saying
\emph{``wealth of information creates a poverty of attention''}.
Furthermore, as humans, we have limited computation capabilities and
increasing number of constrains to develop a single activity, hence, the
formation of \emph{``rules of thumb''} usually takes place instead of
coherent reasoning according to what each state demands. According to
Barber and Odean (\protect\hyperlink{ref-Barber2013}{2013}), the
extension to financial markets stems to the limited devotion to
investing mainly in two fashions: delayed reaction to salient
information and overstated attention to stale information that can lead
to overreaction. As a result, an active agent in crypto marketplace may
face uncertainty and not be able to assess probabilities of events,
accuracy, well-timed choices, the degree of utility, and quality from
some sort of heterogeneous information extracted from sources such as
social media, newspaper, forums, and so on.

Social judgment is intrinsic to cryptocurrency market since the
valuation of any currency is contingent to the extension of the group
that founds it valuable. As with cryptocurrencies, most of the
information technologies exhibit network effects or network
externalities, which is also particularly strong in communication
platforms. Under these scenarios, the strategy is to achieve the
interest of a critical mass of users/investors that yield a higher
market capitalization. Those early adopters (\emph{``Whales''} in
cryptocurrencies' slang) can be positioned and exert market power by
manipulating prices and making profits, this practice normally described
as \emph{``Pump and Dump''}. The objective of boosting prices has as a
mechanism the exposition of exaggerated announcement about the future of
any cryptocurrency, for instance, presumed cryptocurrency's experts
anchor prospects by declaring future increases in prices, narrative
stories of success, any Blockchain's innovative applications in social
media, news, and forums. Once people receive this information, they have
to discern if it is accurate, but prices often react faster, then, it is
strategically rational to follow not only what others do, also movements
of prices. The practice of imitating behavior has been studied in
extension in financial and non-financial markets settings, it has been
named as positive feedback, informational cascades or herding, with some
commonalities and differences which I will try to expose in the section
about crypto-markets speculation.

\hypertarget{the-role-of-news-social-media-and-discussion-forums}{%
\subsection{The role of news, social media and discussion
forums}\label{the-role-of-news-social-media-and-discussion-forums}}

The invention of newspapers permitted a rapid spreading of salient and
not relevant information. Moreover, it also provides as a ploy for the
transmission of hypes with the purpose of capturing reader's attention
towards different issues, being markets one of many of them.

\begin{figure}
\includegraphics[width=1\linewidth]{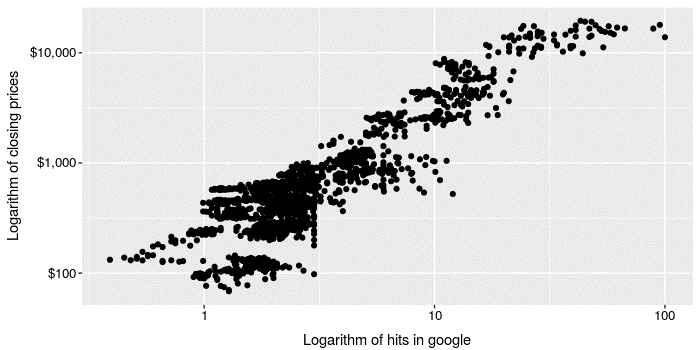} \caption{Google trends hits vs Bitcoin price\label{fig:fig1}}\label{fig:unnamed-chunk-1}
\end{figure}

One important aspect of cryptocurrencies is the fact that prices only
rely purely on market participants' expectations on the future. Hence,
heuristics regarding which cryptocurrency to trust and second which
trade strategy to follow given the information available. Regarding this
point, \autoref{fig:fig1} demonstrates the relevance of people interest
and the correlation with Bitcoin price. Bitcoin is an example of the
limitations of the efficient market hypothesis.

\emph{``A mania involves increases in the prices of real estate or
stocks or a currency or a commodity in the present and near-future that
are not consistent with the prices of the same real estate or stocks in
the distant future''} (Kindleberger, Aliber, and Wiley
\protect\hyperlink{ref-Kindleberger2005}{2005}).

Reflecting a growing recognition of the role of fads and endogenous
market fluctuations, much research has focused in recent years on why
large deviations of market values from fundamentals occur in the first
place and how \emph{``false''} information or fads can be disseminated
in the market. Studying herd behavior\footnote{for a survey, Devenow and
  Welch (\protect\hyperlink{ref-Devenow1996}{1996}) and Bikhchandani and
  Sharma (\protect\hyperlink{ref-Bikhchandani2000}{2000})} has been the
object of considerable effort in recent years for its possible role in
amplifying fads and lead market prices astray from fundamentals.

I suggest that these patterns can be explained by the difficulty of
evaluating a large number of available alternatives for investors to
buy, by investors' tendency to let their attention be directed by
outside sources such as the financial media, by the disposition effect,
and by investors' reluctance to sell short (Barber and Odean
\protect\hyperlink{ref-Barber2001}{2001}).

\hypertarget{positive-feedback-herding-and-informational-cascades}{%
\subsection{Positive feedback, Herding, and Informational
Cascades}\label{positive-feedback-herding-and-informational-cascades}}

None of the theory on behavioral or \emph{``orthodox''} finance has
considered the scenario in which there is no reference to be attached
for. In my view, there are three levels of convictions regarding
positions about markets. The first is associated with the rational
expectations assumption that conveys investors react coherently to
announcements that affect fundamentals. The second degree stems from the
debatable conjecture that prices movements are truly ruled by
fundamentals, in which Shiller
(\protect\hyperlink{ref-Shiller2015}{2015}) has been severely criticized
by showing evidence of an excess of volatility. Finally, we reach a
third-degree exceptionally exposed by crypto-markets, in which by
construction there is no fundamental value, then prices will be
determined in large extension by collective valuation. Noting the
crypto-markets nature, and the compelling evidence regarding human
behavior systematic biases exposed in financial-like markets, which
represent the most evidence supported economic theory, there is a final
question to solve: in the absence of reference points to prices, how do
individuals take decisions in crypto-markets? In a broad context,
comparing the information or digital economy to the industrial, Shapiro
and Varian (\protect\hyperlink{ref-Shapiro1999}{1999}) stated that old
economy differentiates from the new in the substitution of economies of
scale by the economics of networks. That is, in a technological world,
one finds utility as far other people's preferences are aligned. For
instance, a messaging app has as the main purpose of communicating with
a counterpart that can be a group or individual. Nevertheless, if those
whom I want to communicate with does not find the same platform
valuable, makes it worthless for me too. According to the same authors,
in the beginning, it is essential to reach a certain amount of users or
critical mass, and the mechanism to increment the number is driven by a
positive feedback behavior.

It seems coherent to hypothesize that one detonating factor that has
converted Bitcoin into the main cryptocurrency independently for the
fact that it is the first successful cryptocurrency implementation, is a
combination of positive feedback mechanism and self-fulfilling prophecy.
The sociologist Merton (\protect\hyperlink{ref-Merton1948}{1948})
defined a self-fulfilling prophecy as: \emph{``\ldots{}a situation,
evoking a new behavior which makes the originally false conception come
true.''}, translating this situation to our case, it can be interpreted
as those initial opinions which featured digital currencies,
particularly Bitcoin as a milestone of a new era, even though few people
then (probably now too) understand it. Indeed, little of the main
foresight have been fully realized, but reality seems blurry enough to
keep the fad going on. Merton adds: \emph{``For the prophet will cite
actual course of events as proof that he was right from the very
beginning.''} This is potentially related to market value foresight
exposed in social media and forums that declared exaggerated markups
such as 10.000 or 20.000 dollars per BTC that eventually came true.

Anybody has been in a situation where our thoughts and actions are
seemly aligned with what others do. Typical transmission mechanisms are
expressed as word-of-mouth communication, news and social media
exposition, in-place observation, or second degree manifestations such
as market prices (Grossman and Stiglitz
\protect\hyperlink{ref-Grossman1976}{1976}). One important feature
herding or behavioral convergence is that it entails a coordination
mechanism, it can be a social learning heuristic by observing other
decision-makers or coordination based on some signal such as price
movements(Devenow and Welch \protect\hyperlink{ref-Devenow1996}{1996}).
Moreover, the among the range of situations where it has been reviewed
we mentioned investor trading, managerial investment, financing choices,
analyst following and forecasts, market prices, market regulation, bank
runs, bubbles, and welfare (Brunnermeier and Oehmke
\protect\hyperlink{ref-Brunnermeier2013}{2013}; Hirshleifer and Hong
Teoh \protect\hyperlink{ref-Hirshleifer2003}{2003}).

Several attempts have been made to describe crowd behavior in investing
settings, particularly, a seminal article from De Long et al.
(\protect\hyperlink{ref-DeLong1990}{1990}) reintroduced the
\emph{``noise''} concept\footnote{formerly attributed to Black
  (\protect\hyperlink{ref-Black1986}{1986}) who defined as the
  \emph{``opposite of information''}}. According to De Long et al.
(\protect\hyperlink{ref-DeLong1990}{1990}) perspective, noise trader
represents the irrational alter ego of the sophisticated investors, an
investor which misperceive expected returns and generate beliefs and
heuristics to buy and sell following a simple feedback rule to form
insights about market dynamics (Lux
\protect\hyperlink{ref-Lux1995}{1995}). Among the results exposed by De
Long et al. (\protect\hyperlink{ref-DeLong1990}{1990}) I highlight that
under the assumption of unpredictability in opinions and beliefs bared
by noise traders, they can earn higher returns than sophisticated
investors even though they distort prices, generating anomalies such as
an excess of volatility and mean reversion.

From the behavioral economics perspective, the literature on crowd
behavior is called herding. It is defined a decision-making approach
characterized by mimicking actions of others, concretely, Kumar and
Goyal (\protect\hyperlink{ref-Kumar2015}{2015}) defines it as a
\emph{``situation wherein rational people start behaving irrationally by
imitating the judgments of others while making decisions.''}, it is also
defined as any behavior similarity/dissimilarity conveyed by the
interaction of individuals (Hirshleifer and Hong Teoh
\protect\hyperlink{ref-Hirshleifer2003}{2003}). According to Graham
(\protect\hyperlink{ref-Graham1999}{1999}) the herding literature can be
organized into four distinct categories:\footnote{Some other authors
  include payoff externalities (network externalities) models that show
  that the payoffs to an agent adopting an action increases in the
  number of other agents adopting the same action Devenow and Welch
  (\protect\hyperlink{ref-Devenow1996}{1996}). However, they have little
  relation with this study, further literature can be viewed in
  Hirshleifer and Hong Teoh
  (\protect\hyperlink{ref-Hirshleifer2003}{2003})} informational
cascades, reputational herding, investigative herding and empirical
herding, conversely. An informational cascade is described as a process
that stems when someone (optimally) choose to ignore her private
information and instead jump to the bandwagon by mimicking the actions
of individuals who acted previously (Banerjee
\protect\hyperlink{ref-Banerjee1992}{1992}; Bikhchandani, Hirshleifer,
and Welch \protect\hyperlink{ref-Bikhchandani1992}{1992}; Graham
\protect\hyperlink{ref-Graham1999}{1999}). In Bayesian reasoning
context, it is the process of updating posteriors by gradually shrinking
prior's weight as new and supposedly strong information is presented in
a sequential manner. Or in other words, cascades assumes that private
signal (prior) likelihood ratios are unbounded. As a result, it is
likely that individuals in further chain of events will also fall into
mimicking due to the overwhelming nature of the mass beliefs, providing
no useful information for latter observers-actioners.

Among the most related and relevant theoretical contributions we have
Banerjee (\protect\hyperlink{ref-Banerjee1992}{1992}) who found that
decision rules chosen by optimizing individuals will be characterized by
herd behavior. Bikhchandani, Hirshleifer, and Welch
(\protect\hyperlink{ref-Bikhchandani1992}{1992}) provided proofs that
informational cascades could explain conformity, fads, fashions, booms
and crashes. Along the same lines with informational cascades,
Scharfstein and Stein (\protect\hyperlink{ref-Scharfstein1990}{1990})
stated that in individual investment environments (and reputational
herding), managers usually disregard private information by adopting a
follow-the-crowd strategy which is an inefficient behavior from market
perspective, albeit, this situation is rational from their individual
standpoint. Similarly, Welch (\protect\hyperlink{ref-Welch1992}{1992})
results show in IPO settings where shares are sold sequentially, latter
investors based their buying decisions on previous actions, and by
extension forming cascades. Among the causes of herding we can mention
limits of attention exposed before can also increase the probability of
herding or cascade due to the difficulty to accurately process
information (Hirshleifer and Hong Teoh
\protect\hyperlink{ref-Hirshleifer2003}{2003}). It is import to
highlight that `rational herding (maximizing the individual market
participant's utility) could involve the creation of negative
externalities (Banerjee \protect\hyperlink{ref-Banerjee1992}{1992};
Bikhchandani, Hirshleifer, and Welch
\protect\hyperlink{ref-Bikhchandani1992}{1992}).

On the empirical side Welch (\protect\hyperlink{ref-Welch2000}{2000})
found that analysts herd in their stock recommendations from data about
buy and sell, exposing significant positive correlation between adjacent
analysts. Additionally, Welch showed that analyst's elections are
correlated with the prevailing forecast and asymmetry towards a tendency
to herd under the existence of optimism or positive news, concluding
that this situation can create fragility and further crashes. Those
results are aligned with a famous phrase in Keynes
(\protect\hyperlink{ref-keynes1936general}{1936}) which says:
\emph{``Worldly wisdom teaches that it is better for reputation to fail
conventionally than to succeed unconventionally''}. Stracca
(\protect\hyperlink{ref-Stracca2004}{2004}) explains that several
factors may reinforce a tendency to herding, including reputation in a
principal-agent context if the performance of the portfolio manager (the
agent) is costly to monitor, and the fact that compensation is often
computed comparing with other investors performance, pushing risk-averse
traders to conform to the ``average'' assessment of the market{[}Cambiar
esta redacción{]}.

After describing some of the evidence, we can draw understandings about
Bitcoin price formation, given cryptocurrency markets idiosyncrasy
formation of pure beliefs and fuzzy expectations. Particularly, herding
in crypto-markets could stem through price coordination mechanism that
is it can be (errors are implicit) the most efficient social learning
model. This is described by empirical herding category, which has been
studying investors' behavior when they do a momentum-following or
positive feedback investment that is, taking decision based on price
patterns (Sornette \protect\hyperlink{ref-Sornette2003}{2003}).

\hypertarget{strategies-and-price-bubbles}{%
\subsection{Strategies and price
bubbles}\label{strategies-and-price-bubbles}}

The last two chapters were focused on describing common biases and
heuristics to crypto-investors respectively. We have determined that
individuals in crypto-markets have several incentives to chase the
action, and rely their investments by observing prices and using them as
a coordination mechanism due to the lack of salient information or
fundamental news. This scenario seemly relates to the Internet Bubble
(also known as dot-com, or Y2K) when companies like Amazon, Ebay, and
Yahoo! emerged. It was characterized by an over-expectation of future
profits, as a product of recent rises prices for internet related firms,
investor were eager to invest in companies that were associated with
e-commerce, fiber optics, servers, chips, software, improved hardware,
telecommunications or any prefix that could sound as part of the
\emph{``new economy''} Kindleberger, Aliber, and Wiley
(\protect\hyperlink{ref-Kindleberger2005}{2005}). The bubble was
characterized by a rapid increasing NASDAQ Composite index, coming from
1300 in 1996 to 5400 only three years later\footnote{It is relevant to
  highlight that during December 1996 Alan Greenspan (chairman of the
  Federal Reserve Board) coined the famous concept of \emph{``Irrational
  Exuberance''} to illustrate the effect of psychology in stock markets.}.
According to Ofek and Richardson
(\protect\hyperlink{ref-Ofek2001}{2001}) rational explanations had
little power to explain what happened, since internet stock prices were
significantly deviated from their underlying fundamentals and volatility
of prices were out/bounded expressing over-optimistic sentiment, lack of
caution, and the panic of \emph{``not being part''} among the investors.
Particularly, on the last element, cryptocurrency slang has a special
acronym to express this behavior, it is known as \#FOMO or \emph{``Fear
of Missing Out''}, that is, the anxiety of not get into the market when
an unexpected event unchain a rapid valorization of a certain digital
coin. Another example about investors' irrationality was Black Monday,
the crash that took place in October 29 of 1987, on this matter Shiller
(\protect\hyperlink{ref-Shiller1987}{1987}) expressed that nothing
seemed to be different during those days among the investors whom he
surveyed, perhaps, a perception that the market was overpriced, he also
emphasized in the existence of large price movements without any news
breaks, which is not consistent with the EMH which had been criticized
for other authors (De Bondt and Thaler
\protect\hyperlink{ref-DeBondt1985}{1985}). Shiller insisted in stating
that crashes\footnote{Consistent with the argument that noisy
  participants can affect markets in a non-transitory fashion.} seems to
be determined endogenously by investors, either by reaction to others'
actions or manifestations expressed in prices (from here devised the
concept of positive feedback trading). Moreover, some investors conveyed
they rely on \emph{``gut feeling''} as their forecasting method (in
contraposition to fundamental or technical analysis). One of the main
aspects to consider in such scenarios is the impact of speculative price
movements, particularly Kindleberger, Aliber, and Wiley
(\protect\hyperlink{ref-Kindleberger2005}{2005}) stated that:

\emph{``The insiders destabilize by driving the price up and up and then
sell at or near the top to the outsiders. The losses of the outsiders
necessarily are equal to the gains of the insiders. {[}\ldots{}{]} But
the professional insiders initially destabilize by exaggerating the
upswings and the downswings; these insiders follow the mantra that the
`trend is my friend.' At one stage, these investors were known as `tape
watchers;' more recently they have been called `momentum investors.' The
outsider amateurs who buy high and sell low are the victims of euphoria
that affects them late in the day. After they lose, they go back to
their normal occupations to save for another splurge five or ten years
in the future.''}

On one side there are those who think markets are rational and
efficient, are explaining deviations as an exceptional movement from
fundamental value. The other side is composed by those who believe
psychological behavior as the main driver. The way investor believes
they act as they are more intelligent as the average investor in the
market, hence having a big chance to take out the money safe and the
sound is described by Read (\protect\hyperlink{ref-Read2012}{2012}) who
mentioned:

\emph{``Since the crash is not a certain deterministic outcome of the
bubble, it remains rational for investors to remain in the market
provided they are compensated by a higher rate of growth of the bubble
for taking the risk of a crash, because there is a finite probability of
`landing smoothly', that is, of attaining the end of the bubble without
crash.''}

In this work, it has been hypothesized that when prices in
cryptocurrency goes up or down can be attributed in great to herding or
positive feedback reaction to past price changes. The parallelism to the
insiders can be attributed to the whales in crypto-markets seemly behave
as insiders or private informed investors which as Kindleberger, Aliber,
and Wiley (\protect\hyperlink{ref-Kindleberger2005}{2005}) states,
manipulate price's movements as to destabilize the market by
artificially creating exaggerated successive upswings and prices
decrease to make the less informed to buy high and sell low, making them
only victims of the euphoria (Shiller
\protect\hyperlink{ref-Shiller1999}{1999},
\protect\hyperlink{ref-Shiller2015}{2015}; Shleifer
\protect\hyperlink{ref-Shleifer2004}{2004}).

\newpage

\hypertarget{methodology}{%
\section{Methodology}\label{methodology}}

To date few methods have been developed to test for empirical herding
under prices settings. In the literature review section it has been
mentioned that direct observation on investors' actions is the best
approach to test for herding, since the coordination mechanism and the
potential tilting towards the social convention is transparent from the
flow of information dynamics within individuals. Nonetheless, in
cryptocurrency market this is almost impossible due to its privacy,
hence this study will follow prices as coordination mechanism.

This limitation is not unique, in several financial settings analyzing
stocks or exchanges rates almost impossible to get information of market
participants. Given that herding cannot be measured directly from
financial markets, the literature has developed different proxies for
detecting herding behavior based on return's regression tests. This
study employs the methodology present in (Chang, Cheng, and Khorana
\protect\hyperlink{ref-Chang2000}{2000}), which is an improvement from
original methodology offered by Christie and Huang
(\protect\hyperlink{ref-Christie1995}{1995}). Christie and Huang
(\protect\hyperlink{ref-Christie1995}{1995}) suggested the use of
Cross-Sectional Standard Deviation of returns (CSSD) to identify herding
behavior in financial markets, it is defined as:

\begin{equation}
\label{eq:eq1}
CSSD_t=\sqrt{\frac{\sum_{i=1}^n \left( R_{i,t}-\bar{R}_{m,t} \right)^2}{N-1}}
\end{equation}

where \(R_i\) is the observed stock return on a firm \(i\) (in our study
it is described as \(c\) as presented in the data section) at time
\(t\), while \(CSSD_T\) is the cross sectional average of the returns in
the aggregated portfolio at time \(t\). The implicit indication of the
CSSD is that it quantifies the average proximity of individuals' returns
to the mean, by extension, CSSD will always be equal or above zero,
where a value tied to the lowest bound expresses a situation when all
returns flow in harmony while a deviation from the zero mark represents
dispersion. According to Christie and Huang
(\protect\hyperlink{ref-Christie1995}{1995}) it is possible to test for
herding under market stress (large upswings and downswings) events by
exploiting investors' tendency to overturn their private beliefs in
favor of the market consensus. This conclusion stems from a rational the
Capital Asset Pricing Model\footnote{The CAPM relates risk of an
  investment and the expected returns given a market benchmark, which in
  stock market settings is for many cases the S\&P500, deriving a
  measure of sensibility and asset is in comparison to the movements of
  the market. In this study Iit has been established it as a base line
  to denote rationality in cryptocurrency markets.} (CAPM) which
predicts that the dispersion will increase with the absolute value of
the market return since individual assets differ in their sensitivity to
the market return. On the other side, if herding exists, individual
returns will not differ greatly from the market results. Christie and
Huang (\protect\hyperlink{ref-Christie1995}{1995}) empirical tests is
estimated as the \autoref{eq:eq2}:

\begin{equation}
\label{eq:eq2}
CSSD_t=\alpha+\beta^LD_t^L+\beta^UD_t^U+\varepsilon_t
\end{equation}

where:

\(D_t^L=1\) if market return on day lies in the extreme lower tail of
the distribution, or zero otherwise\\
\(D_t^U=1\) if market return on day lies in the extreme upper tail of
the distribution, or zero otherwise

This model was developed to capture differences in investor behavior
upon extreme upswings or downswings in comparison to what it is expected
to be \emph{``normal''}, expressed as the 90\% or 98\% percent of the
distribution. Nonetheless, this methodology have two main drawbacks,
firstly, it is too sensitive to outliers and secondly, it is completely
arbitrary what is considered as \emph{``extreme''} since the 1\% and 5\%
rule might not fit good for all distributions. Consequently this study
will followed an version to Christie and Huang's model proposed by
Chang, Cheng, and Khorana
(\protect\hyperlink{ref-Chang2000}{2000})\footnote{Chang, Cheng, and
  Khorana (\protect\hyperlink{ref-Chang2000}{2000}) stated that Christie
  and Huang (\protect\hyperlink{ref-Christie1995}{1995}) approach
  \emph{``requires a far greater magnitude of non-linearity in the
  return dispersion and mean return relationship for evidence of herding
  than suggested by rational asset pricing models''}.} which is based on
the Cross-Sectional Absolute Deviations defined as:

\begin{equation}
\label{eq:eq3}
CSAD_t=\frac{1}{N}|R_{i,t}-\bar{R}_{m,t}|
\end{equation}

\begin{figure}[h]

{\centering \includegraphics[width=1\linewidth]{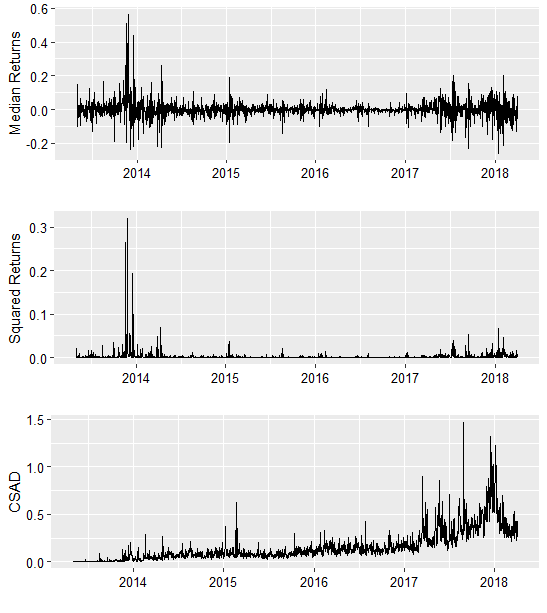} 

}

\caption{Actual and squared cryptocurrency’ market median returns and CSAD\label{fig:fig2}}\label{fig:unnamed-chunk-2}
\end{figure}

The CSAD is a measure of dispersion that takes the absolute difference
between the individuals return and the average market returns, which
makes it far less sensitive to return's outliers than quadratic one.
\autoref{fig:fig2} illustrates the CSAD measure for the full sample, in
which it is noticeable a structural break in the first quarter of 2017
characterized by a level disruption and a higher degree of dispersion.
Chang, Cheng, and Khorana (\protect\hyperlink{ref-Chang2000}{2000})
demonstrated \emph{``that rational asset pricing models predict not only
that equity return dispersions are an increasing function of the market
return but also that the relation is linear''}. Moreover, they rely on
the following intuition: \emph{``if market participants tend to follow
aggregate market behavior and ignore their own priors during periods of
large average price movements, then the linear and increasing relation
between dispersion and market return will no longer hold. Instead, the
relation can become non-linearly increasing or even
decreasing\ldots{}''} This model has been recently employed by several
papers, for instance, Arjoon and Shekhar
(\protect\hyperlink{ref-Arjoon2017}{2017}) examined herding in the
context of frontier market, Chiang and Zheng
(\protect\hyperlink{ref-Chiang2010}{2010}) found herding behavior in
advanced stock markets, Demirer, Lee, and Lien
(\protect\hyperlink{ref-Demirer2015}{2015}) empirically tested for
herding commodity financialization settings and Balcilar, Balcilar,
Demirer, and Hammoudeh (\protect\hyperlink{ref-Balcilar2013}{2013}) who
studied for herding in Gulf Arab stock markets. Following the line of
the aforementioned papers, this study starting with a reference model
specified as:

\begin{equation}
\label{eq:eq4}
CSAD_t=\gamma_0+\gamma_1|R_{m,t}|+\gamma_2R_{m,t}^2+\varepsilon_t
\end{equation}

The model exposed in \autoref{eq:eq4} aims to detect significant
dispersion of returns during markets stress. Hence, a statistically
significant negative coefficient of i.e.~indicates that herding is
likely to be occurring, whereas a significant positive implies a
presence of adverse herding. On identification of herding Kabir and
Shakur (\protect\hyperlink{ref-Kabir2018}{2018}) highlights what Wohar
and Gebka (\protect\hyperlink{ref-Wohar2013}{2013}) stated about a
possible situation when investors \emph{``overemphasize their own view
or focus on views dominant among subset of actors (who may herd jointly
moving in and out of positions) excessively ignoring market information,
it results in increased dispersion in returns across assets leading to
adverse herding''}. It is important to clarify that as many other author
that had been studying herding behavior (Arjoon and Shekhar
\protect\hyperlink{ref-Arjoon2017}{2017}; Chiang and Zheng
\protect\hyperlink{ref-Chiang2010}{2010}; Economou, Katsikas, and
Vickers \protect\hyperlink{ref-Economou2016}{2016}) this model employs
Newey and West (\protect\hyperlink{ref-Newey1987}{1987}) clever solution
to account for heteroscedasticity and autocorrelation consistent
standard errors in regression coefficients, besides the inclusion of
lagged dependent variables ( to guarantee that effects are not a
consequence of autocorrelation dynamics.

Since herding varies across time flows, it would be interesting to
determine whether there are specific periods when herding behavior is
manifested and when it is not, hence, this study will include a regime
Markov Switching (MS) approach to identify regimes in which herding is
exhibited. A MS regression is a useful method to express adjustments
which are more pronounced in high frequency data, moreover, it offers
advantages to reveal patterns commonly hidden in data such as
non-linearity. Regarding the amount of regimes, the definition is not a
straightforward task, on this matter, Psaradakis and Spagnolo
(\protect\hyperlink{ref-Psaradakis2003}{2003}) states that dynamic
models with parameters that are allowed to depend on the state of a
hidden Markov chain have become a popular tool for modelling time series
subject to changes in regimes, nonetheless, the determination of an
adequate number of states to characterize the observed data it is not
conclusive. The MS models offer an advantage over the linear models due
to their ability to reveal patterns beyond traditional stylized facts,
which only nonlinear models can generate. In Psaradakis and Spagnolo
(\protect\hyperlink{ref-Psaradakis2003}{2003}) view, a rule of thumb for
autoregressive models based on AIC values do provide a good instrument
to choose the correct state dimension.

\hypertarget{data}{%
\subsection{Data}\label{data}}

According to the site coinmarketcap.com up to April 2018 there were 1564
different cryptocurrencies available in the market, nonetheless, this
study dampens the sample to the first 100 leading ones which in
aggregated terms account for nearly 96\% of total cryptocurrency's
market capitalization. Getting information about all cryptocurrencies
prices, market capitalization and descriptions is not a completely easy
task. The easy way would be to buy information on specialized websites
that sell datasets, however, it has been scraped the website
www.coinmarketcap.com. The original data includes open, close, highest
and lowest prices, besides its current market capitalization given a day
for each CC. Since crypto-markets are relative new, it is easy to deduce
that non all 100 original presented CC have the same starting dates,
particularly, the two with more observations (1801) are Bitcoin and
Litecoin which extends from April 29, 2013 and ends as all the rest in
April 3, 2018.

Measuring herding intensity by analyzing prices demands to work with
returns, thus I determined the daily return of each c cryptocurrency
arithmetic as follows (\autoref{eq:eq5}):

\begin{equation}
\label{eq:eq5}
R_{c,t}=\frac{P_{c,t}-P_{c,t-1}}{P_{c,t-1}}
\end{equation}

Where \(R\) denotes the price returns of cryptocurrency \(c\) on day
\(t\), and \(CP\) is the closing price of cryptocurrency.
\autoref{tab:table1} reports some descriptive statistics of returns for
the first 50 cryptocurrencies. One of the most iconic features of
cryptocurrencies is the existence of large deviations from the mean,
this volatility is exposed by the existence of long tail distribution
for most of the sample CC this study considered. For instance, taking in
count the subsample seen in table 1, the \emph{``grand''} average return
is 1.3\%, while the average median is -0.1\%, as a result is not
surprise to find a third moment average of 3.7.

\begin{landscape}

\begin{longtable}[]{@{}llllllllllll@{}}
\caption{Summary statistics of returns for the top 50
cryptocurrencies\label{tab:table1}}\tabularnewline
\toprule
\begin{minipage}[b]{0.11\columnwidth}\raggedright
Cryptocurrency\strut
\end{minipage} & \begin{minipage}[b]{0.04\columnwidth}\raggedright
Obs.\strut
\end{minipage} & \begin{minipage}[b]{0.05\columnwidth}\raggedright
Mean\strut
\end{minipage} & \begin{minipage}[b]{0.04\columnwidth}\raggedright
S.D.\strut
\end{minipage} & \begin{minipage}[b]{0.05\columnwidth}\raggedright
Min\strut
\end{minipage} & \begin{minipage}[b]{0.05\columnwidth}\raggedright
Q25\strut
\end{minipage} & \begin{minipage}[b]{0.05\columnwidth}\raggedright
Median\strut
\end{minipage} & \begin{minipage}[b]{0.05\columnwidth}\raggedright
Q75\strut
\end{minipage} & \begin{minipage}[b]{0.05\columnwidth}\raggedright
Max\strut
\end{minipage} & \begin{minipage}[b]{0.06\columnwidth}\raggedright
Skewness\strut
\end{minipage} & \begin{minipage}[b]{0.06\columnwidth}\raggedright
Kurtosis\strut
\end{minipage} & \begin{minipage}[b]{0.08\columnwidth}\raggedright
Jarque Bera\strut
\end{minipage}\tabularnewline
\midrule
\endfirsthead
\toprule
\begin{minipage}[b]{0.11\columnwidth}\raggedright
Cryptocurrency\strut
\end{minipage} & \begin{minipage}[b]{0.04\columnwidth}\raggedright
Obs.\strut
\end{minipage} & \begin{minipage}[b]{0.05\columnwidth}\raggedright
Mean\strut
\end{minipage} & \begin{minipage}[b]{0.04\columnwidth}\raggedright
S.D.\strut
\end{minipage} & \begin{minipage}[b]{0.05\columnwidth}\raggedright
Min\strut
\end{minipage} & \begin{minipage}[b]{0.05\columnwidth}\raggedright
Q25\strut
\end{minipage} & \begin{minipage}[b]{0.05\columnwidth}\raggedright
Median\strut
\end{minipage} & \begin{minipage}[b]{0.05\columnwidth}\raggedright
Q75\strut
\end{minipage} & \begin{minipage}[b]{0.05\columnwidth}\raggedright
Max\strut
\end{minipage} & \begin{minipage}[b]{0.06\columnwidth}\raggedright
Skewness\strut
\end{minipage} & \begin{minipage}[b]{0.06\columnwidth}\raggedright
Kurtosis\strut
\end{minipage} & \begin{minipage}[b]{0.08\columnwidth}\raggedright
Jarque Bera\strut
\end{minipage}\tabularnewline
\midrule
\endhead
\begin{minipage}[t]{0.11\columnwidth}\raggedright
0x\strut
\end{minipage} & \begin{minipage}[t]{0.04\columnwidth}\raggedright
230\strut
\end{minipage} & \begin{minipage}[t]{0.05\columnwidth}\raggedright
0.010\strut
\end{minipage} & \begin{minipage}[t]{0.04\columnwidth}\raggedright
0.112\strut
\end{minipage} & \begin{minipage}[t]{0.05\columnwidth}\raggedright
-0.290\strut
\end{minipage} & \begin{minipage}[t]{0.05\columnwidth}\raggedright
-0.054\strut
\end{minipage} & \begin{minipage}[t]{0.05\columnwidth}\raggedright
-0.006\strut
\end{minipage} & \begin{minipage}[t]{0.05\columnwidth}\raggedright
-0.054\strut
\end{minipage} & \begin{minipage}[t]{0.05\columnwidth}\raggedright
0.632\strut
\end{minipage} & \begin{minipage}[t]{0.06\columnwidth}\raggedright
1.328\strut
\end{minipage} & \begin{minipage}[t]{0.06\columnwidth}\raggedright
7.9\strut
\end{minipage} & \begin{minipage}[t]{0.08\columnwidth}\raggedright
296.0\strut
\end{minipage}\tabularnewline
\begin{minipage}[t]{0.11\columnwidth}\raggedright
Aeternity\strut
\end{minipage} & \begin{minipage}[t]{0.04\columnwidth}\raggedright
306\strut
\end{minipage} & \begin{minipage}[t]{0.05\columnwidth}\raggedright
0.017\strut
\end{minipage} & \begin{minipage}[t]{0.04\columnwidth}\raggedright
0.168\strut
\end{minipage} & \begin{minipage}[t]{0.05\columnwidth}\raggedright
-0.701\strut
\end{minipage} & \begin{minipage}[t]{0.05\columnwidth}\raggedright
-0.056\strut
\end{minipage} & \begin{minipage}[t]{0.05\columnwidth}\raggedright
0.007\strut
\end{minipage} & \begin{minipage}[t]{0.05\columnwidth}\raggedright
-0.056\strut
\end{minipage} & \begin{minipage}[t]{0.05\columnwidth}\raggedright
0.875\strut
\end{minipage} & \begin{minipage}[t]{0.06\columnwidth}\raggedright
0.751\strut
\end{minipage} & \begin{minipage}[t]{0.06\columnwidth}\raggedright
9.6\strut
\end{minipage} & \begin{minipage}[t]{0.08\columnwidth}\raggedright
577.7\strut
\end{minipage}\tabularnewline
\bottomrule
\end{longtable}

\end{landscape}

\newpage

\newpage

\hypertarget{empirical-results}{%
\section{Empirical results}\label{empirical-results}}

\hypertarget{estimates-of-herding-behavior}{%
\subsection{Estimates of herding
behavior}\label{estimates-of-herding-behavior}}

In this section, it has been presented the estimates for the models in
the methodology. The first model is the standard (linear) herding model
which is common in the literature, and we will herein refer to it as the
static model because it has constant parameters. The second model is the
Markov-switching (nonlinear) model which accommodates herding over
multiple regimes. A Markovian switching herding model can be illustrated
as \autoref{eq:eq5}:

\begin{equation}
\label{eq:eq6}
\begin{matrix}
CSAD_{t, 1}=\gamma_{1,0}+\gamma_{1,1}|R_{m,1}|+\gamma_{1,2} R_{m,t}^2+\gamma_{1,k}CSAD_{t-k}+\varepsilon_{t, s} && \varepsilon_{t, s}=N(0,\sigma_1^2) && S_t=1 \\
CSAD_{t, 2}=\gamma_{2,0}+\gamma_{2,1}|R_{m,2}|+\gamma_{2,2} R_{m,t}^2+\gamma_{2,k}CSAD_{t-k}+\varepsilon_{t, s} && \varepsilon_{t, s}=N(0,\sigma_2^2) && S_t=2 \\
\vdots &&  \vdots && \vdots \\
CSAD_{t, s}=\gamma_{s,0}+\gamma_{s,1}|R_{m,s}|+\gamma_{s,2} R_{m,t}^2+\gamma_{s,k}CSAD_{t-k}+\varepsilon_{t, s} && \varepsilon_{t, s}=N(0,\sigma_s^2) && S_t=s
\end{matrix}
\end{equation}

Where \(p\) is defined as the transition probability of the Markovian
chain that can be illustrated as , hence, is the probability of being in
regime \(S\) at time \(t\) given that the in \(S_t\) the regime is equal
to \(j\). Therefore, know the model will be able to identify when
exhibits herding or not, besides different magnitudes this behavior.

\begin{table}[h]
\centering
\caption{Regression estimates of herding behavior on the full sample}
\label{tab:tab2}
\begin{threeparttable}
\begin{tabular}{lllllllll}
\hline
\multicolumn{1}{c}{\multirow{2}{*}{Coef.}} & \multicolumn{2}{c}{\multirow{2}{*}{Static}} & \multicolumn{6}{c}{Regimes} \\
& \multicolumn{2}{c}{} & \multicolumn{2}{c}{1} & \multicolumn{2}{c}{2} & \multicolumn{2}{c}{3} \\ \hline
Intercept & 0.005* & 1.883 & 0.025*** & 3.138 & 0.025*** & 3.138 & 0.025*** & 3.138 \\
$\|R_{m,t}\|$ & 0.240*** & 3.17 & 1.955*** & 4.5 & -0.385*** & -5.066 & 0.581*** & 5.189 \\
$R_{m,t}^2$ & -0.27 & -1.079 & -9.979** & -2.479 & 1.645*** & 8.989 & -0.902*** & -3.685 \\
$CSAD_{t-1}$ & 0.430*** & 18.998 & 0.371*** & 6.29 & 0.336*** & 8.403 & 0.401*** & 9.423 \\
$CSAD_{t-2}$ & 0.220*** & 9.098 & 0.193** & 2.573 & 0.136** & 2.016 & 0.184*** & 3.652 \\
$CSAD_{t-3}$ & 0.277*** & 12.227 & 0.280*** & 3.855 & 0.177*** & 6.669 & 0.227*** & 8.678 \\ \hline
$R^2$ & \multicolumn{2}{c}{0.79} & \multicolumn{2}{c}{0.50} & \multicolumn{2}{c}{0.76} &  \multicolumn{2}{c}{0.79} \\ \hline
$AIC$ & \multicolumn{2}{c}{-4128.8} &  \multicolumn{6}{c}{-6004.4} \\ \hline
\end{tabular}
    \begin{tablenotes}
      \small
      \item This table presents the estimated coefficients of equation 4: $CSAD_t=\gamma_0+\gamma_{s,1}|R_{i,t}|+\gamma_{s,2}R_{m,t}^2+\gamma_{s,k}CSAD_{t-k}+\varepsilon_t$ for the existence of herding. In this specification the intercept is static, that is, it does not change across regimes, while other variables not. The numbers in second row are t-statistics, whereas ***, ** and * stands for significance at 1
    \end{tablenotes}
\end{threeparttable}
\end{table}

\autoref{tab:tab2} reports the estimates for static model and a three
regime switching models according to the specification seen in equation
3. The coefficients are were estimated using Newey and West
(\protect\hyperlink{ref-Newey1987}{1987}) methodology, to achieve
heteroscedastic and autocorrelation consistent standard error estimates
for the full sample, that is, the main 100 cryptocurrencies according
their market capitalization. As I have explained before, under the
assumption that dispersion and the absolute market returns are linearly
related, we must center the attention on the coefficient associated with
, since it captures herding behavior under market stress. In column 3 we
can see that in the static model has a negative sign, nevertheless it is
non-significant. The possible explanation is a high degree of
variability that cancels the effect across the sample, for this matter
it is useful to rely on the Markov Switching estimates that account
dynamics in the parameters. Following the estimates of MS model for the
first regime in column 3 it has been found a significant negative
coefficient of suggesting herding behavior in cryptocurrencies market
the full sample (or \emph{``portfolio''}). Moving to the third regime,
there is statistical evidence in favor of herding, nonetheless the
magnitude is far lower than the first regime exposed above.
Interestingly, there is also a second state where reverse herding is
prominent in the market, seeing at column 3 a coefficient of , an
evidence of market participants behavior characterized by performing
contrary to market consensus, leading to a higher degree of cross
sectional return's dispersion in the cryptocurrency market.

\begin{figure}
\centering
\includegraphics{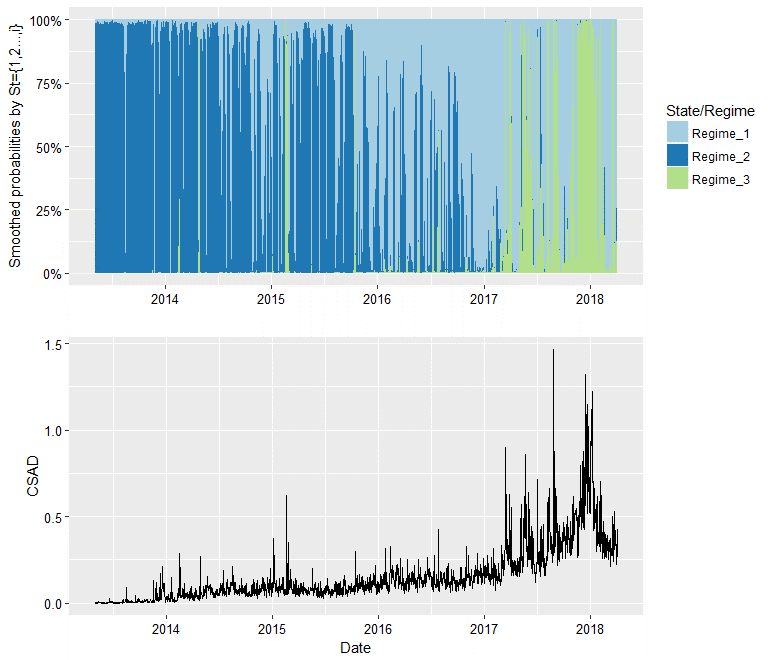}
\caption{Regime switching smoothed probabilities under symmetric herding
behavior for the full sample\label{fig:fig3}}
\end{figure}

The most striking result that stems from the MS model is the
identification of periods where herding behavior has been found
significant. From \autoref{fig:fig3} it can be seen that by far the
greatest amount of time cryptocurrency market exhibits dynamics opposite
to what a rational asset pricing would expect. A further examination of
the graph permits to identify a high probability of regimes 1 and 2 to
fit the data, those two states as explained above had been found to a
strong evidence in favor of herding, prominently on the first regime
(light blue). In addition, the coefficients associated with the absolute
cross sectional returns are significant and positive across all models
(static and MS) suggesting an increasing linear relationship to
dispersion values of CSAD.

\newpage

\hypertarget{estimates-of-herding-behavior-under-asymmetric-market-states}{%
\subsection{Estimates of herding behavior under asymmetric market
states}\label{estimates-of-herding-behavior-under-asymmetric-market-states}}

This investigation began to test the presence of herd behavior in a
sample of the first main cryptocurrencies filtered according their
market capitalization. In the past section, it has been found that
dispersion decreases when extreme returns are present in the market,
nonetheless, it remains to distinguish between the directions in which
returns goes. Regarding this matter, many of the recent empirical
studies coincide in the important to distinguish between herding
behavior under irregular market dynamics, in other words, account for
asymmetric reaction in face of downswings and upswings in the market
returns (Arjoon and Shekhar \protect\hyperlink{ref-Arjoon2017}{2017};
Chiang and Zheng \protect\hyperlink{ref-Chiang2010}{2010}; Demirer and
Kutan \protect\hyperlink{ref-Demirer2006}{2006}; Economou, Katsikas, and
Vickers \protect\hyperlink{ref-Economou2016}{2016}). In order to test
whether crypto-investors react differently on days when the median
returns are positive or negative, it has been created a dummy variable
coded as \autoref{eq:eq7}:

\begin{equation}
\label{eq:eq7}
H(up, down)=\left\{\begin{matrix}
(1-D)R_{m}^2 & if \ R_{m,t} \geq 0  \\ 
DR_{m}^2  & if \ R_{m,t} < 0
\end{matrix}\right.
\end{equation}

\vspace*{.7cm}

Which leads to a new specification given by \autoref{eq:eq8}:

\begin{equation}
\label{eq:eq8}
\begin{matrix}
CSAD_{t, 1}=\gamma_{0, 1}+\gamma_{1,1} D\times \left|R_{m, 1}\right|+\gamma_{2,1} (1-D)\times \left|R_{m, 1}\right|+ \\
\qquad \qquad D\times\gamma_{3,1} R_{m, 1}^2+\gamma_{4,1} (1-D)\times R_{m, 1}^2+\gamma_{4+k,1}CSAD_{t-k,1}+\varepsilon_{t,1} & S_t=1\\
CSAD_{t, 2}=\gamma_{0, 2}+\gamma_{1,2} D\times \left|R_{m, 2}\right|+\gamma_{2,2} (1-D)\times \left|R_{m, 2}\right|+ \\
\qquad \qquad D\times\gamma_{3,2} R_{m, 2}^2+\gamma_{4,2} (1-D)\times R_{m, 2}^2+\gamma_{4+k,2}CSAD_{t-k,2}+\varepsilon_{t,2} & S_t=2\\
 \vdots & \vdots \\
CSAD_{t, s}=\gamma_{0, s}+\gamma_{1,s} D\times \left|R_{m, s}\right|+\gamma_{2,s} (1-D)\times \left|R_{m, s}\right|+ \\
\qquad \qquad D\times\gamma_{3,s} R_{m, s}^2+\gamma_{4,s} (1-D)\times R_{m, s}^2+\gamma_{4+k,s}CSAD_{t-k,s}+\varepsilon_{t,s} & S_t=s\\
\end{matrix}
\end{equation}

\autoref{tab:tab3} reports the regression estimates for herding under
asymmetric conditions for the static and regime switching models as
described in equation 8. In contrast with the model described in
equation 5, this time it has been found that a four regimes fits better
the phenomenon. The static regression estimate of the coefficient
(column 5) confirms the existence of herding when market exhibits
positive returns since parameters leads to enough statistical evidence
in favor of this behavior. On the other side, contrary to our
expectations, there is statistical evidence in favor of reverse herding
under the existence of declining returns , this leads to the conclusion
that crypto-investors do not follow the consensus when market returns
decrease. The fact that when cryptocurrency markets faces extreme
negative returns individuals do not \emph{``flight to safety''}, on the
contrary, it implies that the \emph{``HODL''} strategy is consistent
with the data.

\newgeometry{margin=15mm}
\begin{landscape}

\vspace*{3cm}

\begin{table}[h]
\centering
\caption{Regression estimates of herding behavior on the full sample under asymmetric states}
\label{tab:tab3}
\begin{threeparttable}
\begin{tabular}{lllllllllll}
\hline
\multicolumn{1}{c}{\multirow{2}{*}{Coef.}} & \multicolumn{2}{c}{\multirow{2}{*}{Static}} & \multicolumn{8}{c}{Regimes} \\
& \multicolumn{2}{c}{} & \multicolumn{2}{c}{1} & \multicolumn{2}{c}{2} & \multicolumn{2}{c}{3} & \multicolumn{2}{c}{4} \\ \hline
Intercept & 0.007* & 1.806 & 0.026*** & 10.958 & 0.026*** & 10.958 & 0.026*** & 10.958 & 0.026*** & 10.958 \\
$D\times\|R_{m,t}\|$ & -0.916*** & -4.819 & -0.778*** & -2.885 & -0.017 & -0.037 & 0.480*** & 6.747 & -6.508*** & -3.95 \\
$(1-D)\times\|R_{m,t}\|$ & 0.734*** & 7.384 & 1.251*** & 9.754 & 0.118 & 1.155 & -0.491*** & -24.813 & 4.314*** & 3.662 \\
$D\times\|R_{m,t}^2\|$ & 1.56 & 1.609 & 1.164 & 0.858 & -0.836 & -0.416 & -1.358*** & -3.487 & 25.960** & 2.243 \\
$(1-D)\times\|R_{m,t}^2\|$ & -1.169*** & -4.266 & -2.188*** & -7.63 & 0.376 & 1.303 & 1.372*** & 25.652 & -18.809* & -1.895 \\
$CSAD_{t-1}$ & 0.427*** & 19.363 & 0.374*** & 7.101 & 0.252*** & 7.616 & 0.269*** & 12.634 & 0.372*** & 5.274 \\
$CSAD_{t-2}$ & 0.221*** & 9.36 & 0.259*** & 5.264 & 0.139*** & 5.326 & 0.200*** & 6.636 & 0.201*** & 2.619 \\
$CSAD_{t-3}$ & 0.277*** & 12.534 & 0.241*** & 5.569 & 0.289*** & 14.297 & 0.118*** & 3.99 & 0.220*** & 2.926 \\ \hline
$R^2$ & \multicolumn{2}{c}{0.80} & \multicolumn{2}{c}{0.85} & \multicolumn{2}{c}{0.83} &  \multicolumn{2}{c}{0.91} & \multicolumn{2}{c}{0.57}\\ \hline
$AIC$ & \multicolumn{2}{c}{-4219.5} &  \multicolumn{8}{c}{-6231.4} \\ \hline
\end{tabular}
  \begin{tablenotes}
      \small
      \item This table presents the estimated coefficients of equation 4: $CSAD_t=\gamma_{s,0}+\gamma_{s,1}D|R_{m,t}|+\gamma_{s,2}(1-D)|R_{m,t}|+\gamma_{s,3}D R_{m,t}^2+\gamma_{s,4}(1-D) R_{m,t}^2+\sum_{i=1}^k \gamma_{s,k+4}CSAD_{t-k}+\varepsilon_t$ for the existence of herding. In this specification the intercept is static, that is, it does not change across regimes, while other variables not. The numbers in second row are t-statistics, whereas ***, ** and * stands for significance at 1
    \end{tablenotes}
\end{threeparttable}
\end{table}
\end{landscape}
\restoregeometry

\newpage
\vspace*{2cm}

\begin{figure}[h]

{\centering \includegraphics[width=1\linewidth]{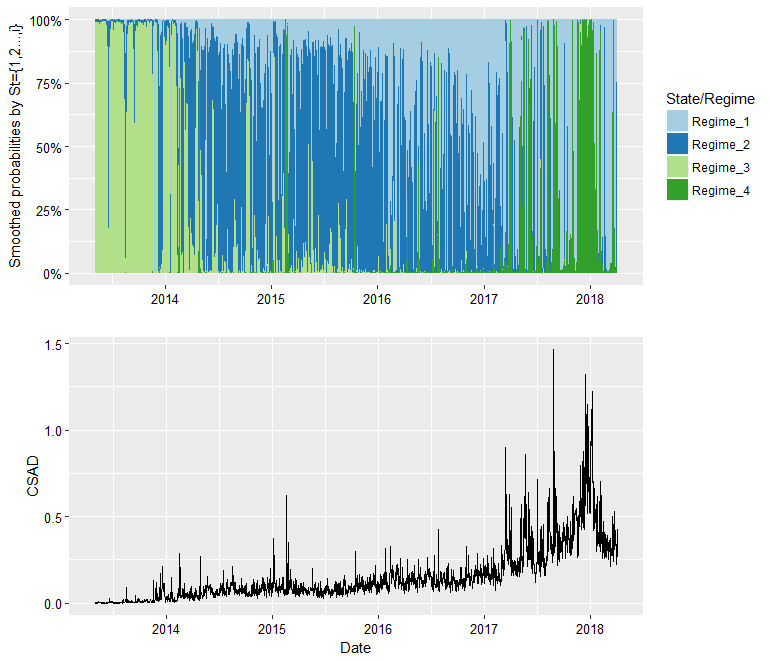} 

}

\caption{Regime switching smoothed probabilities under asymmetric herding behavior for the full sample\label{fig:fig4}}\label{fig:unnamed-chunk-3}
\end{figure}
\newpage

Additionally, it has been estimated the parameters of the MS herding
model under based on equation 8 in order to observe if the static model
fails to capture potential unobserved dynamic structures of herding
behavior over time. The column 4 in table 3 presents the estimates for
herding under extreme declining market settings.

Even though herding behavior under increasing returns situations was
significant at a 1\% level in the static model with , the extension of
the model of MS unveiled the true dynamics inside the data generating
process. Particularly, there is an important difference between the
aforementioned evidence of herding and the one visible in regime 3.
Under a strengthen market situation, is has been found a very strong
coefficient significant at a 10\% threshold level that supports herding.
Looking at \autoref{fig:fig4}, it is clear that during the second
semester of 2017 and the first days of 2018, the cryptocurrency market
exhibited increasing level of cross sectional absolute deviations (or
dispersion), with a substantial reversion from 2018 until the end of the
sample. During that time, the model found that consensus was evident
across investors as the prices started to rise.

\newpage

\hypertarget{conclusion}{%
\section{Conclusion}\label{conclusion}}

This essay was undertaken to evaluate the pertinence of behavioral
finance as a framework to explain price dynamics in crypto-markets
taking as a central point a series of potential biases in decision
making from the investors. Aiming to this objective, it has been
reviewed literature on cognitive biases that has brought evidence of
existence of anomalies, or deviations from what a rational could be
expected in financial related environments. Among the diverse possible
explanations of price movements from a behavioral perspective, the
theory of herding which consists in a situation when individuals ignore
their private information and instead follow the consensus is under
prior consideration a good approach to start the discussion.

The apparent relevance of herding hypothesis to explain price movements
demanded the task of finding an empirical model that led me to study the
phenomena when only prices were the coordination mechanism. The former,
and most relevant methodology to test for herding when only prices are
available is attributed to Christie and Huang
(\protect\hyperlink{ref-Christie1995}{1995}), then is had been improved
for Chang, Cheng, and Khorana (\protect\hyperlink{ref-Chang2000}{2000})
among other authors, this study follow the same line.

The evidence from this study suggests that investors frequently deviated
from the rational asset pricing benchmark, and instead follow the
consensus in market stress situations. This findings have important
implications, first, as I am concerned, this is the first study which
analyzed the price puzzle from herding hypothesis, second, it unveils a
signal that contradicts the circulated \emph{``noise''} exposed in
internet which asserts for the existence of informed people who are not
sensitive to large price movements in cryptomarkets.

\newpage

\hypertarget{references}{%
\section*{References}\label{references}}
\addcontentsline{toc}{section}{References}

\hypertarget{refs}{}
\leavevmode\hypertarget{ref-Alpert1982}{}%
Alpert, Marc, and Howard Raiffa. 1982. ``A progress report on the
training of probability assessors.'' In \emph{Judgment Under
Uncertainty: Heuristics and Biases}, edited by Amos Tversky, Daniel
Kahneman, and Paul Slovic, 294--305. Cambridge: Cambridge University
Press.
\href{https://doi.org/DOI:\%2010.1017/CBO9780511809477.022}{https://doi.org/DOI: 10.1017/CBO9780511809477.022}.

\leavevmode\hypertarget{ref-Arjoon2017}{}%
Arjoon, Vaalmikki, and Chandra Shekhar. 2017. ``Research in
International Business and Finance Dynamic herding analysis in a
frontier market.'' \emph{Research in International Business and Finance}
42. Elsevier B.V.: 496--508.
\url{https://doi.org/10.1016/j.ribaf.2017.01.006}.

\leavevmode\hypertarget{ref-Bachelier1900}{}%
Bachelier, L. 1900. ``Théorie de la spéculation.'' \emph{Annales
Scientifiques de L'École Normale Supérieure} 17: 21--86.
\url{https://doi.org/10.24033/asens.476}.

\leavevmode\hypertarget{ref-Balcilar2013}{}%
Balcilar, Mehmet, Riza Demirer, and Shawkat Hammoudeh. 2013. ``Investor
herds and regime-switching: Evidence from Gulf Arab stock markets.''
\emph{Journal of International Financial Markets, Institutions and
Money} 23 (1). Elsevier B.V.: 295--321.
\url{https://doi.org/10.1016/j.intfin.2012.09.007}.

\leavevmode\hypertarget{ref-Banerjee1992}{}%
Banerjee, Abhijit. 1992. ``A simple model of herd behavior.'' \emph{The
Quaterly Journal of Economics} 32 (3): 378--89.

\leavevmode\hypertarget{ref-Barber2013}{}%
Barber, Brad M., and Terrance Odean. 2013. ``The Behavior of Individual
Investors.'' In \emph{Handbook of Macroeconomics}.
\url{https://doi.org/http://dx.doi.org/10.2139/ssrn.1872211}.

\leavevmode\hypertarget{ref-Barber2001}{}%
Barber, Brad M, and Terrance Odean. 2001. ``Boys Will be Boys: Gender,
Overconfidence, and Common Stock Investment.'' \emph{The Quarterly
Journal of Economics} 116 (1): 261--92.
\url{https://doi.org/10.1093/rfs/hhm079}.

\leavevmode\hypertarget{ref-Barberis2002}{}%
Barberis, Nicholas, and Richard H. Thaler. 2002. ``A Survey of
Behavioral Finance.'' In \emph{Handbook of the Economics of Finance}.
\url{https://doi.org/10.2139/ssrn.327880}.

\leavevmode\hypertarget{ref-Bikhchandani1992}{}%
Bikhchandani, Sushil, David Hirshleifer, and Ivo Welch. 1992. ``A Theory
of Fads, Fashion, Custom, and Cultural Change as Informational
Cascades.'' \emph{Journal of Political Economy} 100 (5): 992--1026.
\url{https://doi.org/10.1086/261849}.

\leavevmode\hypertarget{ref-Bikhchandani2000}{}%
Bikhchandani, Sushil, and Sunil Sharma. 2000. ``Herd Behavior in
Financial Markets: A Review.'' IMF Institute.

\leavevmode\hypertarget{ref-Black1986}{}%
Black, Fischer. 1986. ``Noise.'' \emph{Journal of Finance} 41 (3):
529--43. \url{https://doi.org/10.1111/j.1540-6261.1986.tb04513.x}.

\leavevmode\hypertarget{ref-Brunnermeier2013}{}%
Brunnermeier, Markus, and Martin Oehmke. 2013. \emph{Bubbles, Financial
Crises, and Systemic Risk}. Vol. 2. PB.
\url{https://doi.org/10.1016/B978-0-44-459406-8.00018-4}.

\leavevmode\hypertarget{ref-Chang2000}{}%
Chang, Eric C, Joseph W Cheng, and Ajay Khorana. 2000. ``An examination
of herd behavior in equity markets : An international perspective.''
\emph{Journal of Banking \& Finance} 24: 1651--79.

\leavevmode\hypertarget{ref-Chiang2010}{}%
Chiang, Thomas C., and Dazhi Zheng. 2010. ``An empirical analysis of
herd behavior in global stock markets.'' \emph{Journal of Banking and
Finance} 34 (8). Elsevier B.V.: 1911--21.
\url{https://doi.org/10.1016/j.jbankfin.2009.12.014}.

\leavevmode\hypertarget{ref-Christie1995}{}%
Christie, William G, and Roger D Huang. 1995. ``Following the Pied
Piper: Do Individual Returns Herd around the Market?'' \emph{Financial
Analysts Journal} 51 (4): 31--37.

\leavevmode\hypertarget{ref-Daniel2015}{}%
Daniel, Kent, and David Hirshleifer. 2015. ``Overconfident Investors,
Predictable Returns, and Excessive Trading.'' \emph{The Journal of
Economic Perspectives} 29. American Economic Association: 61--87.
\url{https://doi.org/10.2307/43611011}.

\leavevmode\hypertarget{ref-Daniel1998}{}%
Daniel, Kent, David Hirshleifer, and Avanidhar Subrahmanyam. 1998.
``Investor psychology and investor security market under-and
overreactions.'' \emph{Journal of Finance} 53 (6): 1839--85.
\url{https://doi.org/10.2307/117455}.

\leavevmode\hypertarget{ref-DeBondt1985}{}%
De Bondt, Werner, and Richard H. Thaler. 1985. ``Does the Stock Market
Overreact?'' \emph{The Journal of Finance} 40 (3): 793--805.

\leavevmode\hypertarget{ref-DeLong1990}{}%
De Long, J Bradford, Andrei Shleifer, Lawrence Summers, and Robert
Waldmann. 1990. ``Positive Feedback Investment Strategies and
Destabilizing Rational Speculation.'' \emph{The Journal of Finance} 45
(2): 379--95.

\leavevmode\hypertarget{ref-Demirer2006}{}%
Demirer, Riza, and Ali M. Kutan. 2006. ``Does herding behavior exist in
Chinese stock markets?'' \emph{Journal of International Financial
Markets, Institutions and Money} 16 (2): 123--42.
\url{https://doi.org/10.1016/j.intfin.2005.01.002}.

\leavevmode\hypertarget{ref-Demirer2015}{}%
Demirer, Riza, Hsiang Tai Lee, and Donald Lien. 2015. ``Does the stock
market drive herd behavior in commodity futures markets?''
\emph{International Review of Financial Analysis} 39 (May 2008): 32--44.
\url{https://doi.org/10.1016/j.irfa.2015.02.006}.

\leavevmode\hypertarget{ref-Devenow1996}{}%
Devenow, Andrea, and Ivo Welch. 1996. ``Rational herding in financial
economics.'' \emph{European Economic Review} 40 (3-5): 603--15.
\url{https://doi.org/10.1016/0014-2921(95)00073-9}.

\leavevmode\hypertarget{ref-Dittmar2011}{}%
Dittmar, Jeremiah E. 2011. ``Information Technology and Economic Change:
The impact of the printing press.'' \emph{The Quarterly Journal of
Economics} 126 (3). Oxford University Press: 1133--72.
\url{http://www.jstor.org/stable/23015698}.

\leavevmode\hypertarget{ref-Economou2016}{}%
Economou, Fotini, Epameinondas Katsikas, and Gregory Vickers. 2016.
``Testing for herding in the Athens Stock Exchange during the crisis
period.'' \emph{Finance Research Letters} 18. Elsevier Inc.: 334--41.
\url{https://doi.org/10.1016/j.frl.2016.05.011}.

\leavevmode\hypertarget{ref-Fama1965}{}%
Fama, Eugene F. 1965. ``Random Walks in Stock Market Prices.''
\emph{Financial Analysts Journal} 21 (5). CFA Institute: 55--59.
\url{http://www.jstor.org/stable/4469865}.

\leavevmode\hypertarget{ref-fisher1896appreciation}{}%
Fisher, Irving. 1896. \emph{Appreciation and Interest}. Vol. 11. 4.
American economic association.

\leavevmode\hypertarget{ref-Garber1990}{}%
Garber, Peter M. 1990. ``Famous First Bubbles.'' \emph{Journal of
Economic Perspectives} 4 (2): 35--54.
\url{https://doi.org/10.1257/jep.4.2.35}.

\leavevmode\hypertarget{ref-Graham1999}{}%
Graham, John R. 1999. ``Herding among Investment Newsletters : Theory
and Evidence.'' \emph{The Journal of Finance} 54 (1): 237--68.

\leavevmode\hypertarget{ref-Grossman1976}{}%
Grossman, Sanford J, and Joseph E Stiglitz. 1976. ``Information and
competitive price systems.'' \emph{American Economic Review} 66 (2):
246--53. \url{https://doi.org/10.1126/science.151.3712.867-a}.

\leavevmode\hypertarget{ref-Hirshleifer2003}{}%
Hirshleifer, David, and Siew Hong Teoh. 2003. ``Herd behaviour and
cascading in capital markets: A review and synthesis.'' \emph{European
Financial Management} 9 (1): 25--66.
\url{https://doi.org/10.1111/1468-036X.00207}.

\leavevmode\hypertarget{ref-Kabir2018}{}%
Kabir, M Humayun, and Shamim Shakur. 2018. ``Regime-dependent herding
behavior in Asian and Latin American stock markets.''
\emph{Pacific-Basin Finance Journal} 47 (September 2016). Elsevier:
60--78. \url{https://doi.org/10.1016/j.pacfin.2017.12.002}.

\leavevmode\hypertarget{ref-Kahneman1998}{}%
Kahneman, Daniel, and Mark W Riepe. 1998. ``Aspects of Investor
Psychology.'' \emph{Journal of Portfolio Management}.

\leavevmode\hypertarget{ref-keynes1936general}{}%
Keynes, John Maynard. 1936. \emph{General theory of employment, interest
and money}. Atlantic Publishers \& Dist.

\leavevmode\hypertarget{ref-Kindleberger2005}{}%
Kindleberger, Charles P, Robert Z Aliber, and John Wiley. 2005.
\emph{Manias, Panics, and Crashes}.

\leavevmode\hypertarget{ref-Kumar2015}{}%
Kumar, Satish, and Nisha Goyal. 2015. ``Behavioural biases in investment
decision making -- a systematic literature review.'' \emph{Qualitative
Research in Financial Markets} 7 (1): 88--108.
\url{https://doi.org/10.1108/QRFM-07-2014-0022}.

\leavevmode\hypertarget{ref-Lux1995}{}%
Lux, Thomas. 1995. ``Herd Behaviour, Bubbles and Crashes.'' \emph{The
Economic Journal} 105 (431): 881--96.

\leavevmode\hypertarget{ref-mackay2002extraordinary}{}%
Mackay, Charles. 1852. \emph{Extraordinary popular delusions and the
madness of crowds}. Library of Economics; Liberty.

\leavevmode\hypertarget{ref-Merton1948}{}%
Merton, Robert. 1948. ``The self-fulfilling prophecy.'' \emph{The
Antioch Review} 8 (2): 193--210.
\href{http://www.schirn.de/en/magazine/context/the\%7B/_\%7Dself\%7B/_\%7Dfulfilling\%7B/_\%7Dprophesy/}{http://www.schirn.de/en/magazine/context/the\{\textbackslash{}\_\}self\{\textbackslash{}\_\}fulfilling\{\textbackslash{}\_\}prophesy/}.

\leavevmode\hypertarget{ref-Newey1987}{}%
Newey, Whitney K, and Kenneth D West. 1987. ``A Simple, Positive
Semi-Definite, Heteroskedasticity and Autocorrelation Consistent
Covariance Matrix.'' \emph{Econometrica} 55 (3): 703--8.

\leavevmode\hypertarget{ref-Ofek2001}{}%
Ofek, Eli, and Matthew Richardson. 2001. ``DotCom Mania: A Survey of
Market Efficiency in the Internet Sector.'' \emph{SSRN Electronic
Journal}. \url{https://doi.org/10.2139/ssrn.268311}.

\leavevmode\hypertarget{ref-Poyser2018}{}%
Poyser, Obryan. 2018. ``Exploring the dynamics of Bitcoin's price: a
Bayesian structural time series approach.'' \emph{Eurasian Economic
Review}, no. 0123456789. Springer International Publishing.
\url{https://doi.org/10.1007/s40822-018-0108-2}.

\leavevmode\hypertarget{ref-Psaradakis2003}{}%
Psaradakis, Zacharias, and Nicola Spagnolo. 2003. ``On the determination
of the number of regimes in Markov-Switching Autoregressive Models.''
\emph{Journal of Time Series Analysis} 24 (2).

\leavevmode\hypertarget{ref-Read2012}{}%
Read, Colin. 2012. \emph{The Efficient Market Hypothesists}.
\url{https://doi.org/10.1057/9781137292216}.

\leavevmode\hypertarget{ref-Scharfstein1990}{}%
Scharfstein, David, and Jeremy Stein. 1990. ``Herd Behavior and
Investment.'' \emph{The American Economic Review} 80 (3): 465--79.

\leavevmode\hypertarget{ref-Shapiro1999}{}%
Shapiro, Carl, and Hal R. Varian. 1999. \emph{Information rules}. Vol.
32. 2. Boston: Harvard Business School Press.
\url{https://doi.org/10.1145/776985.776997}.

\leavevmode\hypertarget{ref-Shiller2015}{}%
Shiller, Robert. 2015. \emph{Irrational Exuberance}. Princeton
University Press.
\url{http://www.palgraveconnect.com/doifinder/10.1057/9781137292216}.

\leavevmode\hypertarget{ref-Shiller1987}{}%
Shiller, Robert J. 1987. ``Investor Behaviour in the October 1987 Stock
Market Crash: Survey Evidence.'' \emph{NBER Working Paper}, no. October.
\url{https://doi.org/10.3386/w2446}.

\leavevmode\hypertarget{ref-Shiller1999}{}%
---------. 1999. ``Chapter 20 Human behavior and the efficiency of the
financial system.'' \emph{Handbook of Macroeconomics} 1 (January).
Elsevier: 1305--40. \url{https://doi.org/10.1016/S1574-0048(99)10033-8}.

\leavevmode\hypertarget{ref-Shleifer2004}{}%
Shleifer, Andrei. 2004. \emph{An Introduction to Behavioral Finance}.
Oxford: Oxford University Press.

\leavevmode\hypertarget{ref-simon1982models}{}%
Simon, Herbert Alexander. 1982. \emph{Models of bounded rationality:
Empirically grounded economic reason}. Vol. 3. MIT press.

\leavevmode\hypertarget{ref-Sornette2003}{}%
Sornette, Didier. 2003. \emph{Why Stock Markets Crash}. Vol. 41. 0.

\leavevmode\hypertarget{ref-Stracca2004}{}%
Stracca, Livio. 2004. ``Behavioral finance and asset prices: Where do we
stand?'' \emph{Journal of Economic Psychology} 25 (3): 373--405.
\url{https://doi.org/10.1016/S0167-4870(03)00055-2}.

\leavevmode\hypertarget{ref-Subrahmanyam2008}{}%
Subrahmanyam, Avanidhar. 2008. ``Behavioural finance: A review and
synthesis.'' \emph{European Financial Management} 14 (1): 12--29.
\url{https://doi.org/10.1111/j.1468-036X.2007.00415.x}.

\leavevmode\hypertarget{ref-Svenson1981}{}%
Svenson, Ola. 1981. ``Are we all less risky and more skillful than our
fellow drivers?'' \emph{Acta Psychologica} 47 (2). North-Holland:
143--48. \url{https://doi.org/10.1016/0001-6918(81)90005-6}.

\leavevmode\hypertarget{ref-Welch1992}{}%
Welch, Ivo. 1992. ``Sequential Sales, Learning, and Cascades.''
\emph{The Journal of Finance} 47 (2): 427--65.

\leavevmode\hypertarget{ref-Welch2000}{}%
---------. 2000. ``Herding among security analysts.'' \emph{Journal of
Financial Economics} 58 (3): 369--96.
\url{https://doi.org/10.1016/S0304-405X(00)00076-3}.

\leavevmode\hypertarget{ref-Wohar2013}{}%
Wohar, Mark E, and Bartosz Gebka. 2013. ``International herding: Does it
differ across sectors?'' \emph{Journal of International Financial
Markets, Institutions \& Money} 23: 55--84.
\url{https://doi.org/10.1016/j.intfin.2012.09.003}.

\end{document}